# Frequency Comb-based Wavelength Division Multiplexing and Detection without Wavelength Demultiplexers

Di Che, Zhongdi Peng, Mikael Mazur, Nicolas Fontaine

Nokia Bell Labs, Murray Hill, NJ 07974, United States (*di.che@nokia-bell-labs.com)

**Abstract** *We demonstrate a wavelength division multiplexing (WDM) concept using demultiplexer-free frequency combs at both transmitter and receiver in a 4-$\lambda$ 200-GHz-grid WDM system with flexible symbol rates, aiming to avoid the power-hungry wavelength control on demultiplexers. ©2024 The Author(s)*

**Introduction**

The speed of coherent interface has kept doubling every 3 to 4 years in commercial products since the first generation of 40 Gb/s and reached 1.2 Tb/s in 2023. However, such speed evolution will be more challenging for future generations, considering the symbol rate scaling has started lagging that of the data rate yet the marginal benefit from increasing the modulation order reduces significantly for higher-order formats. While optical networks in the early 1990s used to aggregate multiple IP router ports onto one high-speed wavelength channel, the situation may turn to "inverse multiplexing", where a single interface aggregates more than one wavelength to break the capacity ceiling per wavelength [1]. Such emerging demands on single-interface WDM call for the integration of parallel coherent transceivers onto a single chip. Moreover, coherent optics are expected to be deployed in big volume to a variety of cost-sensitive scenarios like datacenters [2,3], which adds stringent requirements on both integration density and power consumption.

Chip-scale optical frequency combs (OFC) [4-6] have been actively developed over the years for highly parallel WDM transmissions as a promising multi-channel light source. Commonly OFC-based WDM systems cannot live without a wavelength demultiplexer, which separates the comb lines as single-wavelength ones to be individually assigned to the WDM channels as the transmitter (TX) sources or the receiver (RX) local oscillators (LOs). However, implementing demultiplexers on photonic integrated circuits (PICs) is not necessarily a trivial task. The difficulty not only comes from the fabrication of an optical filter bank with high extinction ratio among channels and the potential requirement on wavelength reconfigureability, but the sophisticated on-chip wavelength control with substantial power consumption.

Inspired by the classical optical time-division multiplexing (OTDM) [7] and a variety of related concepts applied at either TX [8-10] or RX [11-14], we demonstrate a comb-based WDM system without using demultiplexers. The basic idea is to apply a series of time delays on the OFC to introduce frequency domain orthogonality, so that the signals modulated on multiple comb lines can be digitally demultiplexed by multi-input-multi-output (MIMO) equalization. The scheme needs neither time synchronization nor optical frequency locking between TX and RX. Compared to traditional comb-based WDM, this demultiplexer-free system does not require any extra hardware, and the only added complexity is from an enlarged MIMO dimension, which is just a small portion of power consumption in coherent digital signal processing (DSP) [3]. Given a fixed WDM grid, each channel can support flexible symbol rates and modulation formats to enable a software reconfigurable system just like in commercial transceivers.

**Principle**

We assume an $N$-channel WDM system with two OFCs for TX and RX, respectively, each having $N$ comb lines. As illustrated in Fig. 1, the demultiplexer in a conventional system is replaced by a colorless power splitter cascaded with a series of time delays $\tau(n)$. Denoting the OFC free spectral resolution (FSR) as $\Delta f$, the delays evenly segment the OFC repetition period ($T = 1/\Delta f$):

$$\tau(n) = (n-1)/N \cdot T, \qquad n = 1,2,\ldots N \qquad (1)$$

According to the time-shifting property of Fourier transform, a time delay results in the phase variations of frequency components proportional to the frequency. Hence, the phase variations of the comb lines for the $n$-th branch ($\vec{\varphi}_n$) due to the delay $\tau(n)$ form an arithmetic sequence spacing at $2\pi\Delta f \cdot \tau(n)$. Using the 1st comb line as reference whose phase variation is assumed to be 0,

$$\vec{\varphi}_n = 2\pi\Delta f \cdot [0 \quad \tau(n) \quad 2\tau(n) \quad \cdots \quad (N-1)\tau(n)] \quad (2)$$

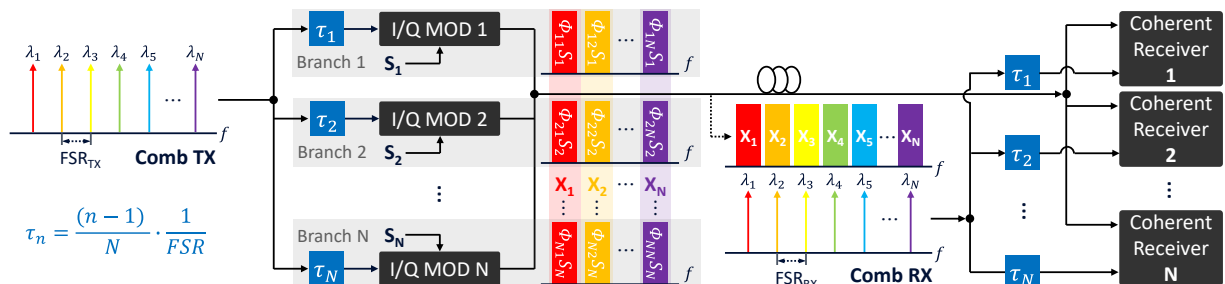

**Fig. 1:** Structure of an OFC-based WDM system ($N$ wavelengths) without using demultiplexers. $f$: horizontal axis of frequency.

We define a matrix that describes the phase variations of the $N$ comb lines (induced by the time delays $\tau$) for all the $N$ TX branches,

$$\boldsymbol{\varphi} = 2\pi\Delta f \cdot \begin{bmatrix} 0 & 0 & 0 & \cdots & 0 \\ 0 & \tau(2) & 2\tau(2) & \cdots & (N-1)\tau(2) \\ \cdots & \cdots & \cdots & \cdots & \cdots \\ 0 & \tau(N) & 2\tau(N) & \cdots & (N-1)\tau(N) \end{bmatrix}$$

$$denote: \boldsymbol{\Phi} = \exp j\boldsymbol{\varphi}$$

$\boldsymbol{\Phi}$ is a unitary matrix, since any of its two columns ($\boldsymbol{\phi}_x$ and $\boldsymbol{\phi}_y$) are orthogonal:

$$\boldsymbol{\phi}'_x \boldsymbol{\phi}_y = \sum_{n=1}^{N} \exp j 2\pi\Delta f(x-y)\tau(n) = 0$$

At TX, the $N$ comb lines of the $n$-th branch are modulated with the same complex signal $S_n$. Due to the preset delay, each branch output forms a phase-shifted series of signals at $N$ wavelengths. We use a power combiner to combine the outputs of all the branches. As shown in Fig. 1, the combined signal at the $n$-th wavelength becomes

$$X_n = \Phi_{1n}S_n + \Phi_{2n}S_n + \cdots + \Phi_{Nn}S_n$$

We denote the input signal as $\vec{S} = [S_1, S_2, \ldots, S_N]^T$ and the WDM signal at the TX output as $\vec{X} = [X_1, X_2, \ldots, X_N]^T$ where $T$ stands for transpose,

$$\vec{X} = \boldsymbol{\Phi}_{TX}^T \cdot \vec{S}$$

At RX, both the WDM signal and the LO comb are power split and sent to $N$ coherent receivers. The lengths of the signal paths are matched while the lengths of the LO paths are adjusted by the time delay series in (1). Inside each receiver, the $N$-line comb LO mixes the $N$-channel WDM signal $\vec{X}$. We choose a receiver bandwidth equal to or lower than the LO comb FSR. Hence, each comb line only mixes the WDM channel closest to its frequency, and the delay-induced phase variations are transferred to the WDM channel during coherent down-conversion. The $n$-th receiver output overlaps the $N$ WDM signals $\vec{X}$ as

$$Y_n = \Phi_{n1}X_1 + \Phi_{n2}X_2 + \cdots + \Phi_{nN}X_N$$

We denote the output of all $N$ coherent receivers as $\vec{Y} = [Y_1, Y_2, \ldots, Y_N]^T$,

$$\vec{Y} = \boldsymbol{\Phi}_{RX} \cdot \vec{X} = \boldsymbol{\Phi}_{RX} \boldsymbol{\Phi}_{TX}^T \cdot \vec{S} \quad (3)$$

The unitary nature of $\boldsymbol{\Phi}_{RX}$ and $\boldsymbol{\Phi}_{TX}^T$ supports the recovery of $\vec{S}$ from $\vec{Y}$ by N×N MIMO equalization without information loss. Because the unitary condition of $\boldsymbol{\Phi}_{RX}\boldsymbol{\Phi}_{TX}^T$ is independent of $\vec{S}$, such MIMO-based demultiplexing supports any modulation formats and flexible symbol rates, so long as the signal bandwidth per wavelength does not exceed FSR$_{TX}$ (if it is equal to FSR$_{TX}$, the WDM signal forms a Superchannel [15]). Though the deviation above only considers phase terms by assuming the same power level for all comb lines, the adaptive MIMO can simultaneously track the power fluctuation among comb lines.

The TX and RX combs do not need to be frequency locked as their frequency offset can be digitally compensated by carrier recovery in receiver DSP. When their FSRs are the same (e.g., using two electro-optic (E/O) combs with precisely controlled radio frequency, RF), the MIMO for RX (i.e., the acquisition of $(\boldsymbol{\Phi}_{RX})^{-1}$) and TX (i.e., $(\boldsymbol{\Phi}_{TX}^T)^{-1}$) can be combined as a unified N×N equalization. If FSR$_{RX}$ is different from the TX WDM grid (i.e., FSR$_{TX}$), the $N$ WDM channels will be down-converted by the $N$ RX comb lines with various frequency offsets. In this case, the MIMO should be separated to two N×N equalizations with a modified digital carrier recovery in between to recover the different frequency offsets.

The above analysis assumes a single polarization (SP) system, but the concept can be extended straightforward to dual polarization (DP) by using DP modulators and coherent receivers, with an expanded MIMO dimension to 2N×2N.

**Experimental setup**
We perform an experiment using the setup in Fig. 2, with a pair of E/O-generated OFCs [16]. The seed lasers are two free-running external cavity lasers (ECL, ~1kHz linewidth) with about 0.01-nm offset. In each comb, the continuous-wave (CW) ECL output is sent to a phase modulator (PM), driven by a 30-dBm power amplifier for a high modulation index. We choose an RF of 33.334 GHz. To obtain an FSR much higher than the RF, we use a wavelength selective switch (WSS) to pick every 1 of 6 comb lines for an FSR of 200 GHz and 4 comb lines in total. Note that for a simple proof of concept, we set the same FSR for the TX and RX combs and use a unified MIMO equalizer (combining TX and RX MIMO in (3)) embedded with a traditional carrier recovery. While the MIMO concept also works with different TX and RX FSRs by modifying the carrier recovery, such demonstrations will be left for our future work.

The TX comb is power split to 4 branches with relative delays (adjusted by 4 variable optical delay lines (VODL)) of 0, 1.25, 2.50 and 3.75 ps, respectively. Each branch has a LiNbO$_3$ SP I/Q modulator (IQM) with a 3-dB bandwidth of ~35 GHz and a smoothly decayed response beyond that. IQM-1 and IQM-2 are modulated by a 4-channel 256-GSa/s arbitrary waveform generator (AWG, Keysight M8199B), and IQM-3 and IQM-4 use a 4-channel 128-GSa/s AWG (M8199A) The two AWGs are synchronized. The 4 IQM outputs are then power combined to form a 4-channel WDM signal. The signal is either back-to-back (b2b) tested or transmitted over 1-km standard single-mode fiber with 17.5-ps/nm/km chromatic dispersion (CD). At RX, the signal is power split to 4 paths with matched lengths. The RX comb is power split to 4 paths serving as 4 LOs with relative delays similar to that for TX. Each pair of signal and LO are fed to a coherent receiver including a 90° optical hybrid, 2 balanced photodiodes, and 2 real-time oscilloscope (RTO) channels at 256 GSa/s. The 8 RTO channels are synchronized for MIMO co-processing. A digital 70-GHz brick-wall lowpass filter is applied on each RTO channel for a unified bandwidth limit. The RX DSP followed routine procedures as listed in Fig. 2, except for the MIMO dimension. This SP 4-channel WDM system requires a 4×4 complex-valued MIMO. However, as the symbol rates are much beyond the bandwidth limit of IQMs, we use an 8×4 real-valued MIMO to compensate for the severer I/Q imbalance at high frequencies. A digital phase locked loop (PLL) is embedded in the MIMO equalizer. We use normalized generalized mutual information (NGMI) for performance evaluation. For each 4-$\lambda$ WDM signal, about 5-million symbols are used to calculate its NGMI.

**Results**

We first verify the effectiveness of MIMO-based wavelength demultiplexing. For each of the 4 TX branches, the time-delayed comb replica is modulated with a 64-QAM signal pulse-shaped by a root-raised cosine (RRC, roll-off factor of 0.01) filter. We generate symbol rates from 80 to 120 GBaud and evaluate their NGMI as shown in Fig. 2(i). For each I/Q signal, NGMI reduces at a higher symbol rate, due to the worse effective number of bits (ENoB) for AWGs and severer bandwidth limit of IQMs. Although the output optical power of the 4 IQMs were equalized by adjusting their drive swings, IQM 3 and 4 perform worse than IQM 1 and 2 due to the worse frequency response and ENoB of AWG M8199A (compared with M8199B). Despite these performance variations, the MIMO equalizer offers a reliable demultiplexing at all symbol rates.

We then fix the symbol rate to 128 GBaud by operating AWG M8199A at 1 sample per symbol (sps). This improves the ENoB performance for M8199A owing to lower peak-to-average power ratio (PAPR) at 1sps (with respect to the RRC-filtered case). We change the entropy of probabilistically shaped (PS) 256-QAM signals from 6.8 to 7.8 bits/symbol and evaluate their NGMI as shown in Fig. 2(ii). To claim a net bit rate (NBR), we choose a concatenated forward error correction (FEC) code (rate 0.7436) [17], with an NGMI threshold (NGMI*) of 0.8105. The maximum entropy that meets the FEC threshold (NGMI*) is 7.4 bits/symbol. This results in an NBR of

$$(7.4 - (1 - 0.7436) \times 8) \times 128 = 684.64 \ (Gb/s)$$

per channel and 2.74 Tb/s in total for 4 channels.

An interesting question is whether CD breaks the orthogonality among TX signals considering it induces wavelength-dependent time delays. A brief answer is no, because CD only introduces a diagonal matrix between $\Phi_{RX}$ and $\Phi_{TX}^T$ in (3) that does not break the unitary condition. To verify this, we transmit the WDM signal over 1-km fiber. The CD-induced delay between two edge WDM frequencies (spacing at 4.8 nm) is ~84 ps, much larger than the delays of TX branches. In Fig. 2(ii), the 1-km transmission shows similar performance to the b2b case, indicating the MIMO demultiplexing is robust to CD.

**Conclusions**

Without demultiplexing the comb lines, we experimentally demonstrate an OFC-based WDM system that supports flexible modulations and symbol rates per wavelength in a fixed grid. It offers a promising chip-scale WDM solution for a single coherent interface with reduced form factor and power consumption on wavelength control.

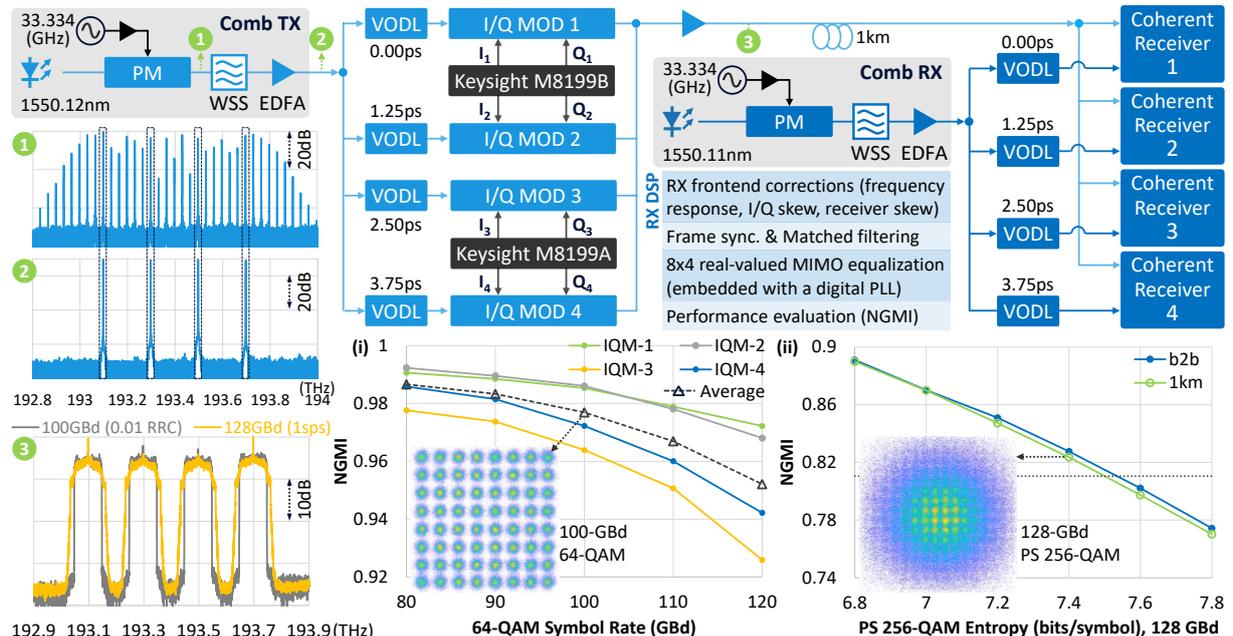

**Fig. 2:** Experiment. 4-channel WDM NGMI results for (i) 64-QAM at various symbol rates; (ii) PS 256-QAM at 128 GBd.